\documentclass[a4paper,11pt]{article}
\pdfoutput=1 

\usepackage{jinstpub} 

\title{Identification and mitigation of memory block timing issue in ITk ABCStar during ASIC production}

\author[f]{B. Ashmanskas,}
\author[a,1]{J. Botte,\note{Corresponding author.}}
\author[a]{J.R. Dandoy,}
\author[d]{J. Dopke,}
\author[f]{N. Dressnandt,}
\author[d]{B.J. Gallop,}
\author[e]{J.J. John,}
\author[f]{P.T. Keener,}
\author[a]{T. Koffas,}
\author[f]{J. Kroll,}
\author[h]{R.P. McGovern,}
\author[f]{F.M. Newcomer,}
\author[a]{B.J. Norman,}
\author[d]{P.W. Phillips,}
\author[d]{C. Sawyer,}
\author[b]{R. Scouten,}
\author[c]{P. Vicente Leitao,}
\author[g]{M. Warren}

\affiliation[a]{Department of Physics, Carleton University, \\
  1125 Colonel By Drive, Ottawa, Ontario, Canada, K1S 5B6}
\affiliation[b]{DA-Integrated,\\
  27 Iber Road, Stittsville, Ontario, Canada, K2S 1E6}
\affiliation[c]{CERN, 1211 Geneva 23, Switzerland}
\affiliation[d]{Particle Physics Department, Rutherford Appleton Laboratories,\\
  Didcot, United Kingdom, OX11 0QX}
\affiliation[e]{Department of Physics, University of Oxford,\\
  Parks Road, Oxford, United Kingdom, OX1 3PU}
\affiliation[f]{Department of Physics and Astronomy, University of Pennsylvania,\\
  209 South 33rd Street, Philadelphia, Pennsylvania, United States of America, 19104-6396}
\affiliation[g]{Department of Physics and Astronomy, University College London,\\
  Gower Street, London, United Kingdom, WC1E 6BT}
\affiliation[h]{Department of Physics and Astronomy, University of Victoria,\\
  3800 Finnerty Road, Victoria, British Columbia, Canada, V8P 5C2}

\emailAdd{james.botte@carleton.ca}

\abstract{The ABCStar is a mixed-signal front-end readout ASIC for the strips sensor portion of the
  ATLAS ITk detector being developed as part of the High-Luminosity LHC upgrade. In pre-production
  testing, a subtle design flaw was uncovered in the ABCStar that was reducing wafer yields in some
  manufactured lots from the expected 90\% to as low as 2\% ``good'' chips. The root cause was
  determined to be a timing issue in the logic synthesized to control previously ``silicon proven''
  memory blocks re-used for this ASIC. The solutions proposed included manufacturing process changes by
  the wafer foundry, changes to the operating parameters for the ABCStar in the detector, and as a last
  resort the possibility that a redesign might be required. The two mitigation efforts were undertaken
  in parallel, with the process modification route a less desirable solution since already manufactured
  wafers would need to be scrapped in favour of the new ones. Based on a knowledge of the existing
  process, and testing done on the worst performing wafers, it was proposed that raising the core
  operating voltage of the ABCStar from 1.20~V to 1.25~V could address the timing issue by sufficiently
  speeding up its transistors. An extensive testing program that included the effects of temperature and
  radiation expected over the lifetime of the ITk detector was conducted --- and thermoelectric studies
  were performed --- to validate that approach. Those tests and studies proved that even the worst
  performing wafers would have yields over 80\% with the 1.25~V core voltage, and neither the modified
  process nor redesign would be required for ensuring reliable operation of the ITk. Based on testing, a
  further timing mitigation was implemented to provide an additional margin of reliability by increasing
  the duty cycle of the clock to the ABCStar. Testing of all ABCStar wafers has been completed and the
  production of the detector modules using these ASICs is now well underway as a result of the efforts
  detailed herein.}


\begin{document}
\maketitle
\flushbottom

\section{Introduction}
\label{sec:intro}

The ATLAS Binary Chip ``Star Architecture'' (ABCStar) is a mixed-signal front-end readout Application
Specific Integrated Circuit (ASIC) for the strips sensor portion of the
ATLAS~\cite{atlas}~\cite{atlas_r3} inner tracker upgrade project (ITk)~\cite{itk}~\cite{itktdr} required
for the High-Luminosity Large Hadron Collider (HL-LHC) upgrade at CERN~\cite{hllhc}. In pre-production
testing, certain digital tests were seen to be failing with unexpected regularity. Specifically, the
output from the physical devices did not match the expected output from a set of \emph{pass/fail}
digital test vectors. The prevalance of the failures seemed to be concentrated in certain batches of the
wafers, on certain wafers within the batches, and more often in certain locations across wafers ---
generally the dice towards the wafer periphery. Furthermore, the failures were seen on dice that passed
all the other, often extensive, tests that exercised both analog and digital functionality.

\begin{figure}[htbp]
\centering 
\includegraphics[width=\textwidth]{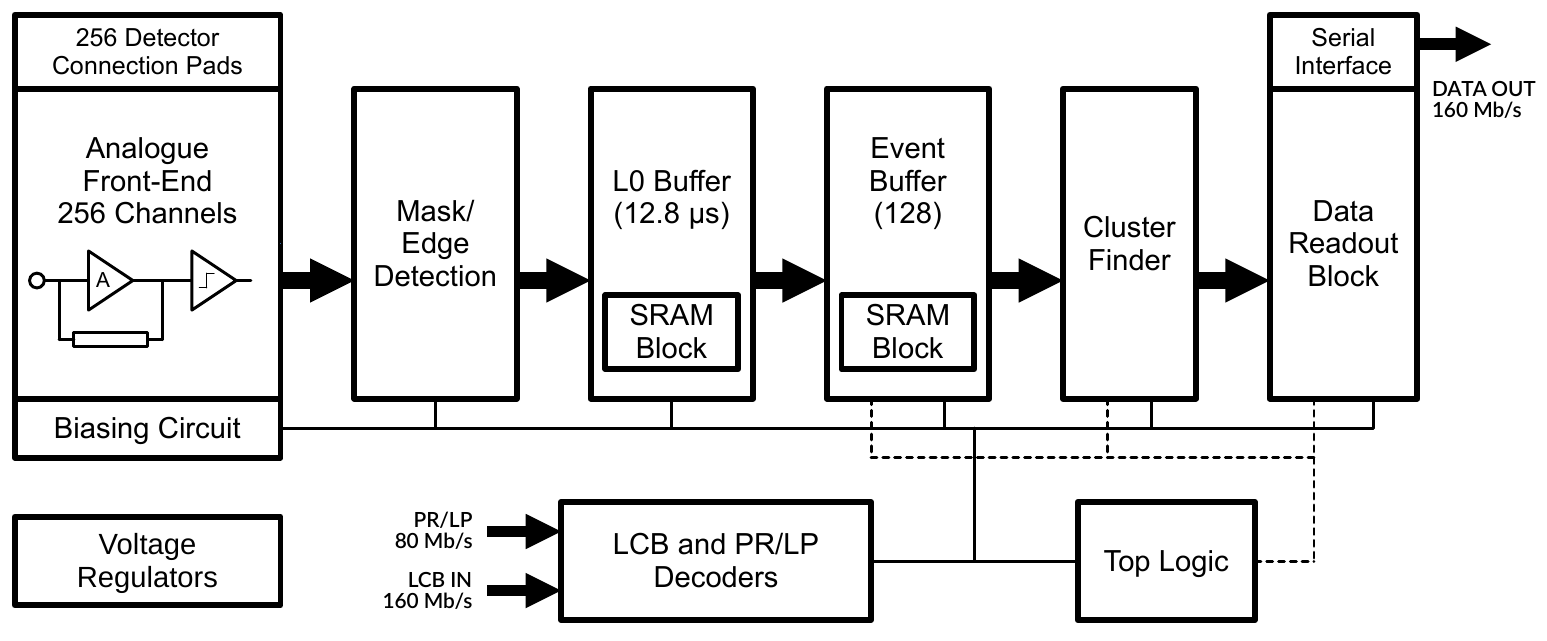}
\caption{\label{fig:abcblock} ABCStarV1 Block Diagram Showing SRAM Blocks}
\end{figure}

Upon examination of the outputs of the physical devices, we were able to see that bit flips in the
output were responsible for the test failures, and that the location of those bit flips was inconsistent
across runs of the same test on the same device. Because the data protocol and packet format on the
output was ``well formed'', but the data within the packets had errors, the the output circuitry of the
chip was not suspect and it became clear the data corruption was originating in Static Random Access
Memory (SRAM) blocks that had a long heritage of reliable operation in other designs, including the
previous revision of the ABCStar itself, and had been considered ``silicon proven''.

An ad hoc multi-disciplinary cross-institution team was convened to further investigate, formulate and
test mitigation strategies, and propose solutions that could be verified in testing. The root cause was
determined to be marginal timing in the circuitry synthesized to interface to the SRAM blocks, not the
SRAM blocks themselves. Proposed solutions included process changes by the ASIC fabrication facility
(the ``foundry''), changes to the operating parameters for the ASIC in situ, and as a last resort the
possiblity that a costly redesign might be required if no other solution proved feasible and reliable.

This article will describe the process of discovering the existence of a problem, determining a clear
root cause of the failures, the options that were explored to address the root cause issue, the
mitigation strategies implemented to address the issue, and the testing of the chosen solution to verify
it was robust and suitable for production and deployment in the ITk detector.

\section{Brief overview of ABCStar ASIC and its testing}
\label{sec:abcstar}

The ABCStar ASIC is comprised of an analog front end with 256 input channels with pulse shaping and a
tunable analog to digital discriminator for each channel \cite{abcstarfe}, digital circuitry to read out
the front end and transmit event data on request over a high speed serial interface to the Hybrid
Controller Chip ``Star Architecture'' (HCCStar) data aggregator and controller ASIC, and various other
analog and digital circuitry blocks. Several ABCStar ASICs are used on each ITk silicon strip module ---
with their input channels wirebonded to a large area silicon strips detector, and their power and
digital signals wirebonded to a polyimide printed circuit board \cite{abc130}. Once the ASICs are
assembled onto a module, should an ASIC fail during further testing, it is very difficult and risky to
replace it and could lead to modules having to be scrapped or to accept performance degredation of such
a module. Furthermore, once the detector is assembled and installed in the ATLAS experiment, it is
expected to have to operate without any intervention for the prescribed lifetime of the ITk detector. As
such, the reliability and quality of the ABCStars used in the assembly is of paramount importance. A
program of testing the ASICs at the wafer level to ensure full functionality --- before the wafers were
diced and used in module assemblies --- was a critical aspect of the ITk strips detector program.

\begin{figure}[htbp]
\centering 
\includegraphics[width=\textwidth]{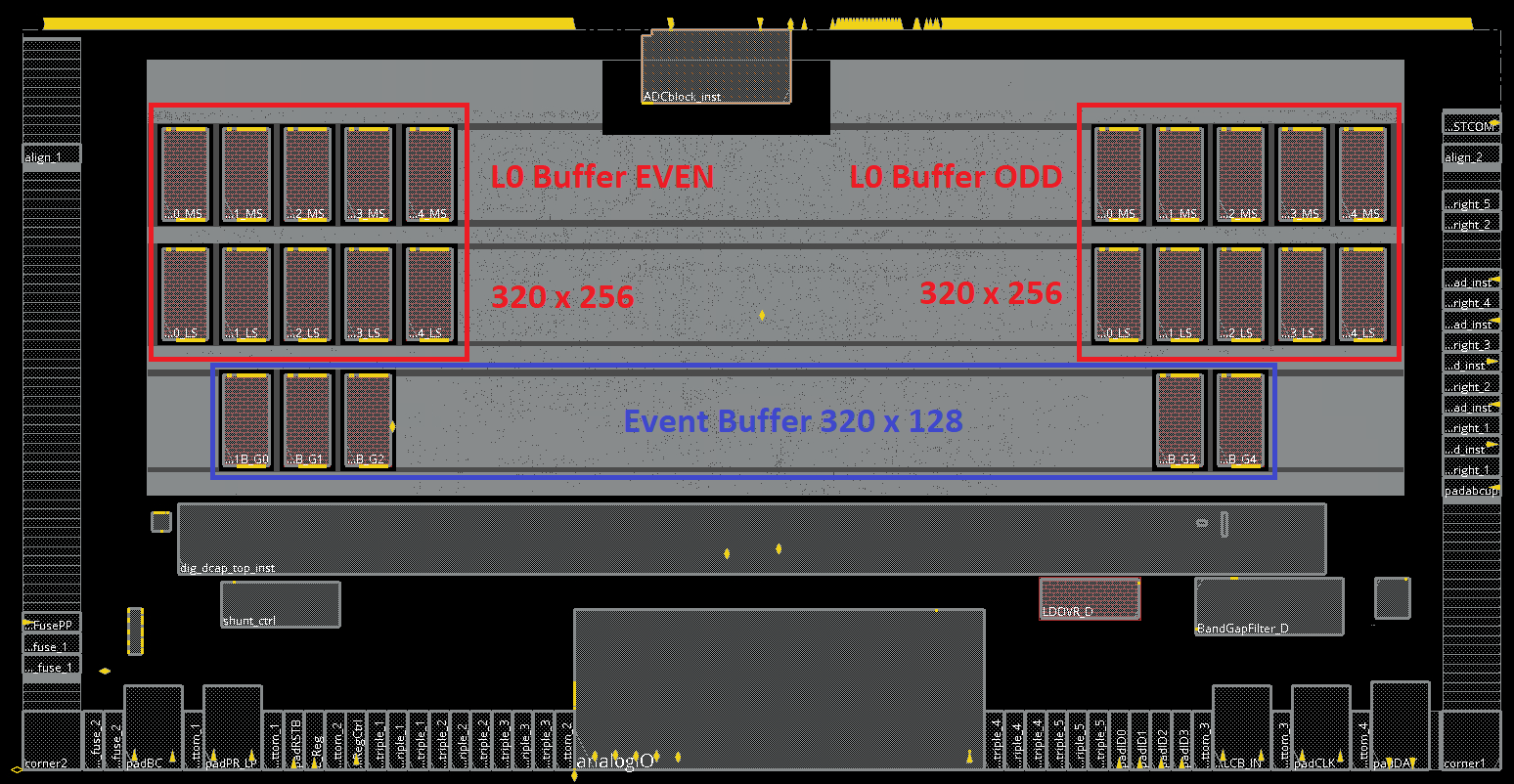}
\caption{\label{fig:floorplan} Location of SRAM blocks on the ABCStar floorplan}
\end{figure}

In order to fully and effectively test ABCStars on a wafer, extensive Built-In Self-Test (BIST) features
were included in the design of the chip. This includes being able to self-inject calibrated analog
pulses directly into the front end of the input channels, the ability to inject digital patterns to
simulate events, and various other features that allow most of the chip to be tested in situ on the
wafer. In referencing Figure \ref{fig:abcblock}, data generally moves from the left to the right of the
diagram with the L0 Buffer and the tagged Event Buffer used to hold data from the front end (the
discriminators and edge detection/mask circuitry) for possible detailed downstream event processing by
the ITk detector's Data Acquisition Systems (DAQ). The Cluster Finder and Readout Block are used to
process event data, then compress and send it efficiently to the data aggregation and analysis stages of
the DAQ through the HCCStar. Of particular importance to this discussion is the Static Random Access
Memory (SRAM) blocks within the L0 Buffer and Event Buffer as this is where the issue with the chip
design occurred.

The L0 Buffer is physically implemented with 64-bit wide by 128-bit deep single-port SRAM blocks
arranged in two arrays five wide by two deep to create a pair of 320-bit wide by 256-bit deep SRAM
memories, see Figure \ref{fig:floorplan}. Events during normal operation are read continuously from the
front end, 256 channels at a time, into one of the two memories comprising the L0 Buffer --- alternating
even or odd --- at 40~MHz. This clock signal is derived from the LHC's fixed frequency beam crossing
clock and is provided to the ABCStars on a module by an HCCStar. The 40~MHz clock drives a 9-bit free
running address counter for the L0 Buffer to sequentially store data into the buffer, where the lowest
bit of the counter selects whether the even or odd memory will be written to; and also an 8-bit free
running but command-resettable Beam Crossing ID (BCID) that is used to detect if synchronization of the
system with the ITk DAQ is ever lost (event data must associated with the LHC collision that generated
it). In addition to the normal operation modes, BIST circuitry can be selected to generate digital
patterns to the edge detection/mask circuitry in place of the discriminator data, see Figure
\ref{fig:abcblock}. Because the 9-bit counter is free running, it wraps around to 0 after it reaches
it's maximum count. As such, the data from an event is available in the L0 Buffer for 12.8~\textmu s
(25~ns times 512 memory locations) before it is overwritten with a new event. 256 of the 320 bits of
each memory word in the L0 Buffer are used for the event data, the value of BCID is triplicated and
stored along with the event for 24 more bits (it is part of the control path and therefore triplicated
where the event data is not), and the remaining bits are unused. The Event Buffer is 320 bits wide by
128 bits deep and is built with the same SRAM blocks as the L0 Buffer. Data from events stored in the L0
Buffer can be transferred to a specified location in the Event Buffer with a synchronous command to the
ABCStar through the HCCStar from the ITk DAQ.

Each ABCStar ASIC (a ``die'' or ``chip'', or plural ``dice'' or ``chips'') is tested in situ on the
wafer it was manufactured on with a wafer-level probing system. Once electrical connection is made with a
die on the wafer, a testing procedure is executed to verify that the chip has sufficient operationality
to allow further testing, followed by extensive exercising and verification of its functionality under a
variety of electrical conditions. Tests are performed in stages to ensure that functionality required
for later, more complex and demanding stages of testing is working as expected before those later tests
are run --- with the test procedure abandoned and the die marked as ``bad'' if any of these tests
fail. For the ABCStar, the major groupings of tests include: basic electrical testing (e.g. prober
continuity with die, checking for shorts and opens, etc.), tuning the core voltages, verifying
functionality of the analog support blocks, verifying functionality of and tuning the analog front end
using its built-in self-test features and gathering statistics about its performance, and running a
series of digital test vectors to stimulate the digital portions of the die. It was in the digital test
vector portion of the procedure where the issue with the chip was detected in pre-production testing.

Each die probed was assigned a category of A, B, or X; where Category A were fully functional and able
to be used in the final detector assembly, B had minor issues with some of the detector input channels
but were otherwise operational, and X consisted of dice that failed basic testing or had too many bad
input channels or other defects. Category B dice were set aside for laboratory testing and in case there
was a shortfall of category A ASICs during final assembly --- at the minor loss of some channels or
performance. Category X dice were used for mechanical, wirebonding, gluing, and other tests that did not
require functional chips. Viable commercial integrated circuit yields --- the number of good die versus
the total number on a wafer --- can vary from 30\% to over 80\% overall based on a chip's design and
surface area (see, for example \cite{yield}); however, due to inherent process variation during
manufacture, differing yields are expected between separately manufactured wafer lots. Based on
experience with similar ASICs developed by CERN using the same 130~nm mixed-signal CMOS process, a yield
of 90\%+ Category A chips was expected for the ABCStar wafers. That yield expectation was used to
estimate the total number of manufactured wafers required to produce enough Category A ABCStars to build
the ITk strips detector, with some spares. With the production testing done, employing the mitigations
described below, the Category A yield is 85.88\% and the Category A+B yield is 89.94\%, which are both
roughly in line with expected yields and demonstrate a good understanding of the manufacturing process
used.

\section{Identification of digital circuitry issue}
\label{sec:issue}

As the wafer-level testing procedures were being developed, functional tests were implemented first to
ensure that the wafer tester, and the test programs being custom developed for the task, could control
and monitor all parts of a device under test (DUT, an ABCStar die on a wafer in this case) in order to
extract and record results under various operating conditions. The tester hardware included programmable
and monitorable power supplies, and analog and high speed digital inputs and outputs. Furthermore, some
analog and digital measurements could be made and read out from the ABCStar with its own telemetry
blocks using commands to the chip over its digital interfaces. Of particular importance for the tests
was gathering statistics on whether the performance of the 256 silicon strip detector input channels of
the ASIC were meeting the required specifications individually and collectively.

Once that phase of test development had been implemented, a number of digital test vectors were
introduced to the testing procedure that had been adapted from the Verilog~\cite{verilog} ASIC
validation and verification test suites used during the ABCStar's design phase. Due to their
comprehensive nature, their inclusion as part of the ASIC test process was seen as a useful final
verification step of a physical die's digital circuitry before it was marked ``good''. While we foresaw
that these digital tests might find subtle and limited manufacturing defects, we expected they would
rarely result in a failure determination. This expectation was based on the robustness of digital
circuitry against process variation due to mature ``design for manufacturability'' techniques included
in the simulation data provided by the foundry (see, for example, \cite{dfm_mc}), and the testing
already done before that point in the procedure. Indeed, after introducing the new digital tests for the
first pre-production wafer lots there were few additional die failures from those tests, and as such the
results did not arouse any particular concern.

\begin{figure}[htbp]
\centering 
\includegraphics[width=\textwidth]{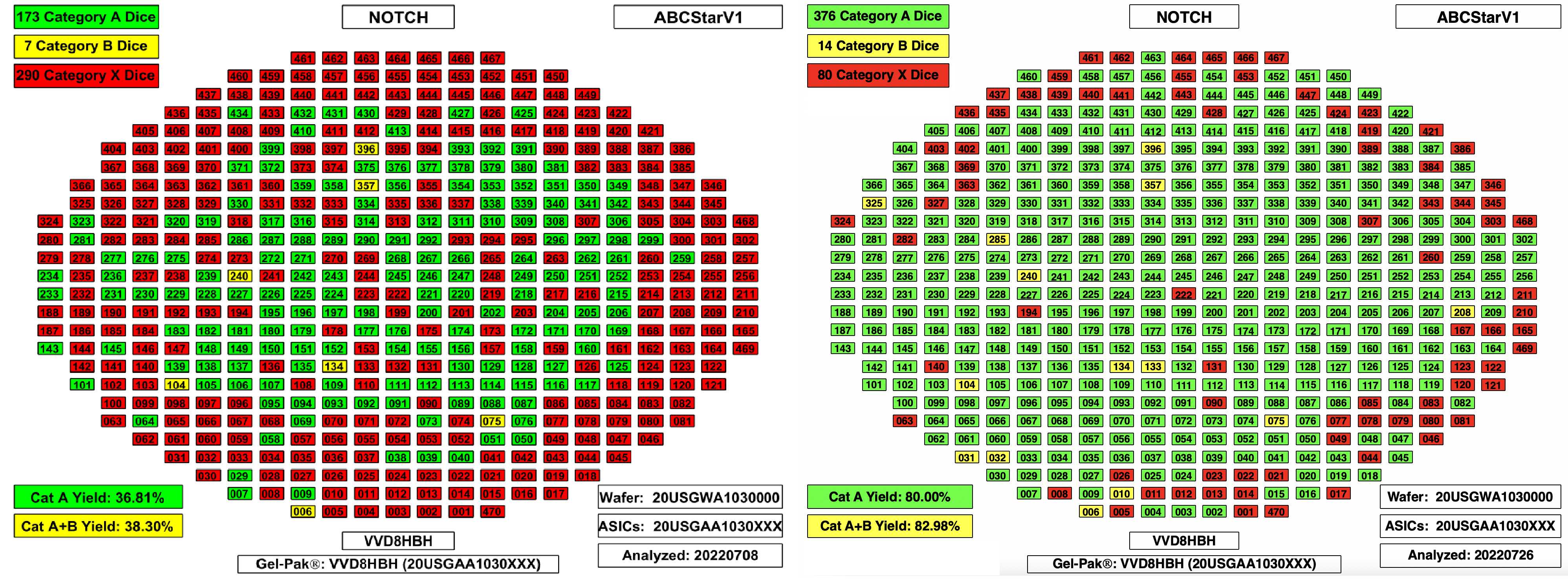}
\caption{\label{fig:badvolt} Example yields from a \emph{bad lot} wafer at 1.20~V (left) and 1.50~V (right)}
\end{figure}

To move into the wafer production testing phase, all tests of the ABCStar needed to be run at the core
voltages that they were going to be used at in the ITk detector. In addition, selected digital tests
were to be run at an even lower voltage to ensure the chips had a generous operating margin. For ease of
implementation during the test development phase, all tests were run at 1.50~V which was supplied by the
wafer tester and not further regulated on chip. Production testing needed to be done at 1.20~V --- the
intended core voltage to be used in the completed detector --- regulated down to that voltage from the
1.50~V supplied by the wafer tester to the ASIC using the on-chip programatically tuneable Low-DropOut
linear regulators (LDOs): one for the digital circuitry and one for the analog circuitry. An additional
test, named P99, a test to check the functionality of the redundancy features on the ASIC through the
datapath (which includes the SRAMs), was to be run at 1.20~V and then again at a 1.10~V core voltage for
the ABCStar. In testing a particularly low yield lot of wafers (the ``bad lot'') in early 2022 at 1.20~V
(and 1.20~V and 1.10~V for the P99 test), serious issues were observed with yields as low as 2\% on one
wafer in that lot. Furthermore, the yields on other wafers in the bad lot would also drop to single
digits if the core voltage was dropped a further 25~mV to 1.175~V, implying significant marginality. The
three tests that showed serious voltage sensitivity were called A02, P03, and P99 per Table
\ref{tab:key_tests}. Another 14 digital tests, some quite complex, never presented any particular issue
at 1.20~V and will not be referred to further here. When the tests were repeated with a core voltage of
1.50~V on wafers in the ``bad lot'', the digital tests that had been failing mostly passed (see Figure
\ref{fig:badvolt}). Rerunning the tests on a ``known good'' wafer lot, we observed failure rates
increased only by a small percentage when run at the lower voltages with the implemented digital
tests. The production tests were to be run at the lower core voltages for both the digital and analog
LDOs, and tests on the bad lot showed that only the digital LDO core voltage output was a determinant of
test performance, and raising or lowering the analog core voltage had no effect on those tests.

\begin{table}[htbp]
\centering
\caption{\label{tab:key_tests} The three primary tests that are sensitive to SRAM timing issue}
\smallskip
\begin{tabular}{|c|p{0.3\columnwidth}|p{0.5\columnwidth}|}
\hline
Test ID&Test Name&Test Description\\
\hline
A02&Readout in BIST Mode&Randomised test that interleaves triggering and readout of hits, various
resets, and reading and writing of chip registers. The chip is configured in a BIST mode such that the
readout pipeline is filled with a time-varying pattern generated from the BCID counter.\\
\hline
P03&Readout with pulse patterns&This test injects varying digital simulated hit pattern pulses and
checks the cluster finder responses over the chip's high-speed serial output.\\
\hline
P99&Test SRAM memory readout&This test uses an internal digital bitmask function to fill the memories
with ones, then varies the memory addressing to read out all bits of the SRAMs to verify the readout
path.\\
\hline
\end{tabular}
\end{table}

\begin{figure}[htbp]
\centering 
\includegraphics[width=\textwidth]{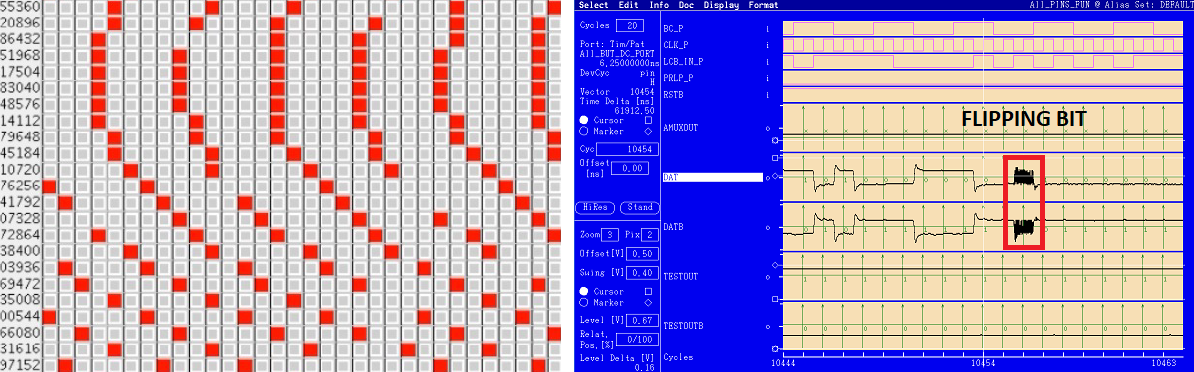}
\caption{\label{fig:bitflips} First indication of data corruption running P99 test at 1.10~V core voltage}
\end{figure}

As stated before, the digital tests are either \emph{pass} or \emph{fail}: if the output from the ASIC
matches exactly what is expected by the design, the test passes; otherwise it fails. Detailed tests were
performed on one particular wafer from the bad batch. Failures of the P99 test at 1.10~V were
intermittent on some dice with the number of errors and location of the errors in the output varying
between different die and runs of the test on the same die, but it always passed at 1.20~V. There were
anywhere from 2 to 900 errors in the output of the test, where an error is a bit that does not match the
expected pattern. Figure \ref{fig:bitflips} on the left shows a portion of one of the first diagnostic
outputs of the problem captured on a wafer probing system showing where the bit flips were occuring in
the output pattern from the DUT for the P99 test running at 1.10~V. Each line is 65536 bits and each
square is 2048 bits within the output, and the number on the left is the number of the first bit in the
line. A red block in a square indicates some number of flipped bits in that block, and the exact number
and which bits specifically were available through a manual data visualizer on the tester. Figure
\ref{fig:bitflips} shows on the right a waveform view of a bit flip on a particular die's differential
serial data output (DAT/DATB) pair. The arrows show the timing at which the output signal from the DUT
is sampled by the wafer tester and the associated 0 or 1 shows the expected output from the DUT. The
test is run repeatedly and if a bit is flipping, the average of the outputs is displayed --- which shows
as the ``scribbled'' bit on the display in this test. The scale on the bottom of the plot indicates the
bit position in the expected pattern. Die that failed the P03 test consistently failed at 1.20~V, with 2
to 26 errors, but passed consistently at 1.50~V. However, it was the A02 test at 1.20~V that caused the
most dice to fail the test suite. On the bad lot, the A02 test was even failing on some die at 1.50~V.
Of particular importance was that the format of the output packets from the chip was always correct, but
it was the simulated but deterministic physics data within the packets' payload that was experiencing
the errors. Suspicion turned quickly to the SRAMs in the buffers that provide data to the cluster finder
to format (see Figure \ref{fig:abcblock}), and that was later proven to be a correct assessment after
further testing. Given the ability of affected dice to pass at higher digital LDO voltages, and for a
sample of wafers from another lot to pass at the lower voltages, it did not imply a logic design error
in the ABCStar but rather an unexpected process-related dependency with one or more internal circuit
blocks.

Efforts were immediately taken to:
\begin{itemize}
  \item try to understand which SRAMs might be the culprit, and why, if that was the source;
  \item to probe wafers from several lots with tests running at various voltages lower than 1.50~V to
    assess the variation in susceptibility of individual dice, within lots, and lot to lot variation;
  \item to gather statistics from more wafers in the bad lot;
  \item and to involve the ASIC design team at CERN.
\end{itemize}

Performance over temperatures and across integrated radiation dose was also noted as being a potential
avenue of investigation as this affects transistor speed along with the core voltage used. Of particular
concern was whether this could impact production yields of the ASICs if the process variation that
resulted in the bad lot could occur in other lots. It should be noted that the foundry that manufactured
the wafers did exercise test structures included in the wafer periphery to ensure the wafers were within
their process variation limits before they were shipped, which they were. Furthermore, those process
variation limits are included in the models provided by the foundry for simulation, and the
functionality of the circuitry is simulated for all corner cases of the process during the design phase
of the ASIC (see e.g. \cite{dfm_corner}). The observed performance did not align with the simulated
performance, and that discrepancy also needed to be investigated. Additionally, when digital tests
failed it tended to be more frequent towards the outer edges of a wafer, which further implied process
variation related susceptibility to these failures between die on a single wafer. A process called spin
coating is used in the manufacturing process of semiconductor devices on a wafer (see
e.g. \cite{spincoat}), and some radial variation of this nature is expected and also included in the
simulation models provided by the foundry.

\begin{table}[htbp]
\centering
\caption{\label{tab:a02_probe} Varying core voltage and clock duty cycle with A02 test on 111 dice}
\smallskip
\begin{tabular}{|c|c|c|}
\hline
LDO Register Setting&Clock Duty Cycle (\%high/\%low)&\# A02 Failures (of 111)\\
\hline
Nominal ($\sim$1.20~V)&50/50&83\\
Nominal ($\sim$1.20~V)&60/40&1\\
Nominal ($\sim$1.20~V)&62.5/37.5&1\\
Nominal + 1 ($\sim$1.175~V)&50/50&100\\
Nominal - 1 ($\sim$1.225~V)&50/50&42\\
Nominal - 2 ($\sim$1.25~V)&50/50&8\\
Nominal - 3 ($\sim$1.275~V)&50/50&1\\
Nominal - 4 ($\sim$1.30~V)&50/50&1\\
\hline
\end{tabular}
\end{table}

\section{Root cause determination}
\label{sec:cause}

The need to pursue multiple avenues of investigation quickly led to the extensive use of the ``shmoo
plot'' \cite{shmoo}. Shmoo plots are a graphical representation of test results across an often
multi-dimensional operational parameter space, e.g. varying core voltage while also varying clock
frequency. In general, a particular test is chosen to execute across the parameter space of interest
based on its sensivitity to the variation in operating conditions. Such tests can simply indicate a pass
or fail result, or can return their own quality parameter that can be displayed graphically on the plot
as a colour variation and/or numerical value. Optimally, such tests need to be performed across a
variety of dice at different locations on a wafer, on different wafers, and on wafers from different
lots to gather robust statistics. Gathering such data provided the investigation team with important
insights into both the root cause of the issue and further suggested potential mitigation strategies
that might be used to avoid a costly redesign. Futhermore, as statistics were gathered from a diverse
population of devices, from both high and low yield wafers for instance, the parameter spaces where
reliable operation was guaranteed across all test-passing devices could be determined and further inform
the root cause investigations.

\begin{figure}[htbp]
\centering 
\includegraphics[width=0.9\textwidth]{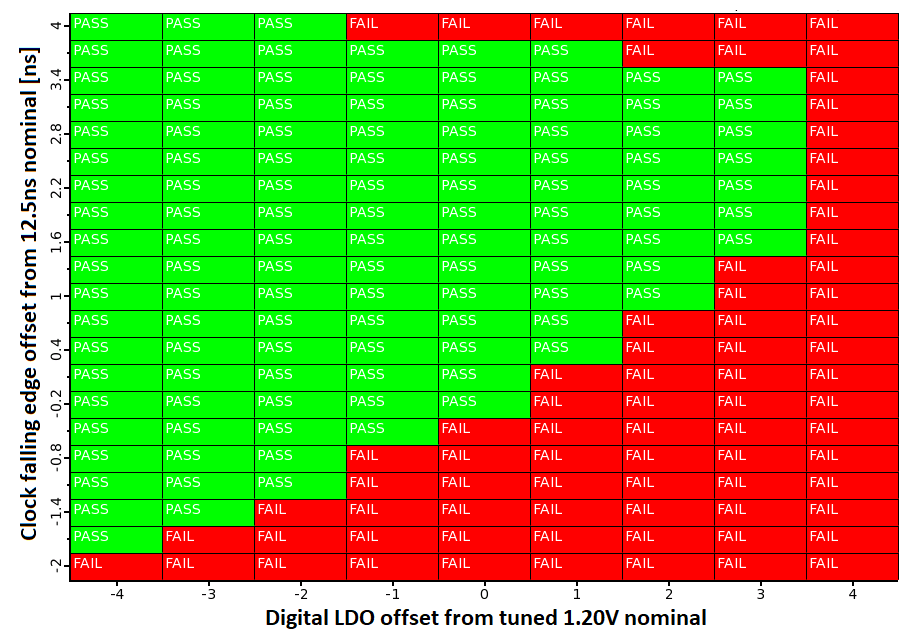}
\caption{\label{fig:a02shmoo} Shmoo plot of one ABCStar die with A02 test: digital LDO versus clock duty cycle}
\end{figure}

In one of the early explorations, the A02 test was chosen as the test probe for a particular wafer in
the ``bad lot'' that had yields of less than 20\%. It was chosen because the A02 test could fail on
chips where the A03 and P99 tests passed, and thus was shown to cause the observed problem more often
than the other tests. It was used are the primary ``probe'' for this behaviour in future testing as
well. Both core voltage and the clock duty cycle were varied on a portion of that wafer: 111 dice, of a
possible 470, were probed with the results shown in Table \ref{tab:a02_probe}. The core voltage on each
chip was tuned through its digital LDO's control register to generate a nominal value of 1.20~V as
measured by the wafer prober's analog monitoring capability. The seven tests shown were run on the die,
possibly incrementing or decrementing the LDO's control register from nominal, then the wafer prober
moved to the next die. The LDO register setting versus the core voltage generated is not a directly
linear relationship, but roughly correspond to the voltages shown in the table. What is clear from the
tests is that the issue is very sensitive to the core voltage, but also to the duty cycle of the clock
for a given core voltage. This was further confirmation that it was a timing issue within the chip
rather than a logic error.

A shmoo plot was performed on a particular die on the same wafer where the digital LDO voltage register
setting was varied by steps on either side of the tuned 1.20~V nominal setting while also varying the
duty cycle. In Figure \ref{fig:a02shmoo}, the duty cycle setting variation is the number of nanoseconds
applied to the nominal 50\% duty cycle position of the falling edge of the clock. For instance, +1.6
resulted in a clock that was high 14.1~ns of the 25~ns cycle for a duty cycle of 56.4\%.  More negative
LDO settings produced higher core voltages than nominal, and more positive were lower core voltages
(roughly $\pm$27.8~mV per step on average).

\begin{figure}[htbp]
\centering 
\includegraphics[width=\textwidth]{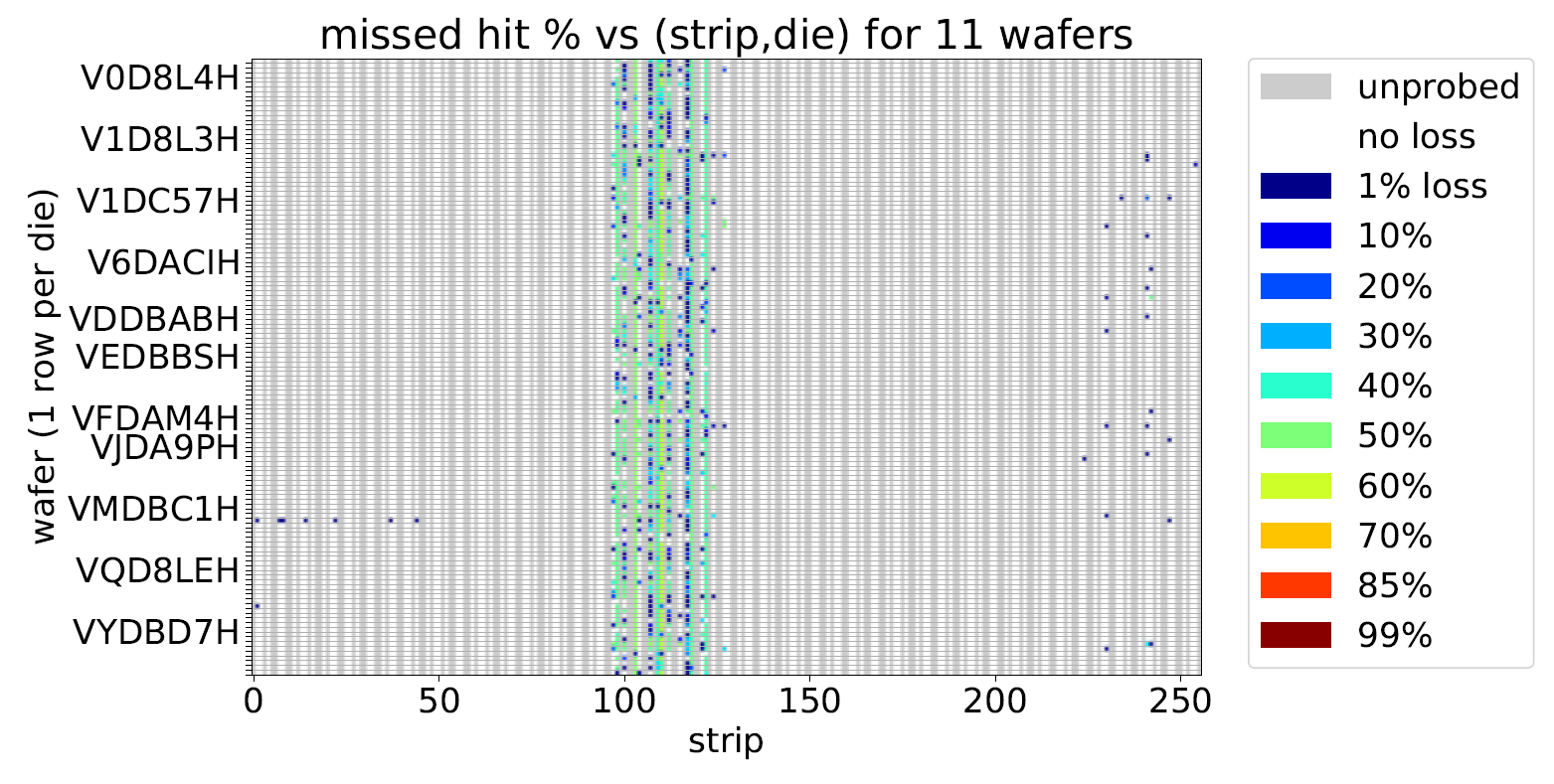}
\caption{\label{fig:srambits} ABCStar data error channel dependency across 11 wafers}
\end{figure}

The SRAM blocks themselves did have a legacy in other ASICs that had been powered at 1.20~V (``silicon
proven'') and the wafers had been manufactured and tested to, and complied with, the specifications used
for the process in simulation, and as such attention was focused on the ``glue logic'' that had been
synthesized around the SRAM blocks. Other experiments at higher core voltages, as discussed previously,
had proven the soundness of the logic design itself. Looking at the channels where these missing hits
occurred, it was possible to determine which specific SRAM array slices were responsible for the errors,
see Figure \ref{fig:srambits}, and further cast suspicion that the ``glue logic'' in the ABCStar that
controlled the read and write accesses to the SRAM blocks had a process-related issue with timing. Given
that access to these single-port SRAM blocks were performed on the falling edge of the 40~MHz clock
(with the access setup beginning on the rising edge of the clock), having the clock remain in the high
state longer (higher duty cycles) allowed more time for logic states to propagate through the ``glue
logic'' before trying to write/read to/from the SRAMs. The tests where increasing the duty cycle reduced
the number of errors seen without changing the core voltage further supported this hypothesis. Speeding
up the transistors through higher core voltages allowed the logic states to propagate faster, and
resulted similarly in more reliable operation.

The A02 test that ran on the wafer testing hardware performed a number of pseudorandom operations on the
ABCStar. The equivalent test that ran on the ASIC during design simulations generated actual random
sequences of operations; however, the wafer tester systems required a fixed ``test vector''. The Verilog
code for the test was run for a period of time, generating random sequences, and the input to the chip
from the simulation code was captured as a string of binary values and used as the A02 test, and thus
was a fixed and deterministic test when run on real hardware so the output from the chip could be tested
against a known good output sequence based on the fixed input test vector. Specifically, the A02 test
made use of a BIST mode called TM11 (Test Mode One One) that used the BCID counter to generate a 0 or 1
based on the XOR of its 8 bits. The 8-bit value of the BCID specified which channel of the edge
detection/mask circuitry that value would be sent to. This simulated various patterns of detector
``hits'', where a ``hit'' in a physical detector is the pulse(s) created in the strips sensor's silicon
when a high-energy particle interacts with it --- which is then detected by a connected ABCStar if it
exceeds the discriminator threshold for channel(s) connected to the strip(s) with pulses. Further
testing with the A02 test showed that the bit flips were due to ``missing hits'' in the SRAMs, where the
deterministic digital data patterns injected into the ABCStar by the A02 test were either not being
stored or read out properly. This issue could have occurred between the front end and the L0 Buffer,
between the L0 Buffer and the Event Buffer, or between the Event Buffer and the Cluster Finder
circuitry; however, the error patterns shown in Figure \ref{fig:srambits} indicated the missing hits
were introduced either in the interface of the mask and edge detection circuit to the L0 Buffer, or the
interface of the L0 Buffer to the Event Buffer. Based on knowledge of the design, and the testing done
on physical chips, it was predicted that the issue could be fully mitigated by increasing the core
voltage to speed up the transistors, by speeding up the transistors in some other way, and/or by
increasing the duty cycle of the clock to remain higher longer to give more time before the falling edge
of the clock for the SRAM glue logic to settle to a stable state before the SRAM blocks were accessed.

The last question that needed to be answered was why the synthesized ``glue logic'' had the timing issue
in the first place. All simulations indicated that timing requirements to access the SRAM blocks would
be satisfied with the generated logic circuitry, but this did not match the real world performance of
the ASICs under expected manufacturing process variations. The source of the design issue was ultimately
traced to the timing models of the SRAM blocks themselves. These blocks had been taken as ``silicon
proven'' and the timing models associated with those blocks had been assumed accurate for the process
being used. This treatment of ``silicon proven'' design blocks is common practice in the ASIC
development community, and the decision to use the SRAM blocks as provided without re-simulation was not
considered a risk. However, in examining the legacy of the SRAM blocks further, the timing models had
been manually generated using previous generations of design tools and were not fully representative of
their actual performance in the real world. When the control logic around them was synthesized, the
tools used for the ABCStar design ensured it met all timing constraints provided to it. Because the
timing model of the SRAM block was ultimately inacurate, even though it had been assumed to be correct
though its proven legacy, the timing issue was present in the correctly manufactured ASIC. This was a
key finding of the investigation and needs to inform future ASIC designers that use any ``silicon
proven'' design blocks without resimulation --- often necessary if such blocks are provided by foundries
or other design houses as opaque intellectual property to be used as provided, again a common practice
in the industry to minimize engineering resource requirements and reduce design risk.

\section{Investigation into mitigation options}
\label{sec:mitigation}

Once the root cause was identified, it was possible to formulate a test strategy to determine if
anything could be done to mitigate the issue. Two primary options were explored in hopes of avoiding a
costly re-design of the ABCStar: changes to the operating conditions of the ASIC in situ, and
modifications to the wafer production process. The first option was, of course, preferred but required
extensive testing to validate any approach that looked promising before it could be approved. At the
same time, the foundry was engaged and made several suggestions on ways to speed up transistors on the
ASICs through manufacturing process changes.

Both options were explored in parallel with testing on existing ABCStar chips over all environmental and
operating conditions they would be subjected to during operation in the ITk detector, and the production
of several experimental wafers that modified transistor behaviour was initiated. The main issue with the
experimental wafer route was that nearly half of the planned wafers at the time had already been
manufactured with the initial process and could not be modified, and representative test chips using the
modified transistors would also have to go through extensive validation to make sure there were no
unexpected side effects. If a modified process resulted in the only viable go-forward strategy, the
existing wafers would likely need to be scrapped and replaced at great expense. However, this was still
preferable than having to re-design the ASIC as that would require the scrapping of all tooling that had
been produced for the original design --- the non-recurring engineering and process effort needed to
introduce and validate a new design at a foundry is expensive and time consuming.

\subsection{Investigation of possible mitigations for already manufactured ABCStars}
\label{sec:nomodasic}

At the outset of the investigation, determination was made that all wafers that had been supplied did
meet the quality criteria of the foundry in their tests and were within the acceptable range of
manufacturing process variation --- it was a design issue on our end that made some chips susceptible to
that variation. This, and the test results we had already gathered, gave us confidence that we properly
understood the technology and process we were using. As such, we knew a higher core voltage or lower
temperature would result in faster transistors which in turn would likely result in a reliable SRAM
block with no timing issues. Additionally, due to the way the SRAM ``glue logic'' was designed and
implemented, if the clock's duty cycle to the ABCStar was higher (i.e. if the logic high portion of the
clock signal was longer than the logic low portion) it would allow that circuitry more time to settle
and not require faster transistors to operate reliably. Since the effects of radiation were more subtle,
tests were required to provide experimental data for the investigation.

\begin{figure}[htbp]
\centering 
\includegraphics[width=\textwidth]{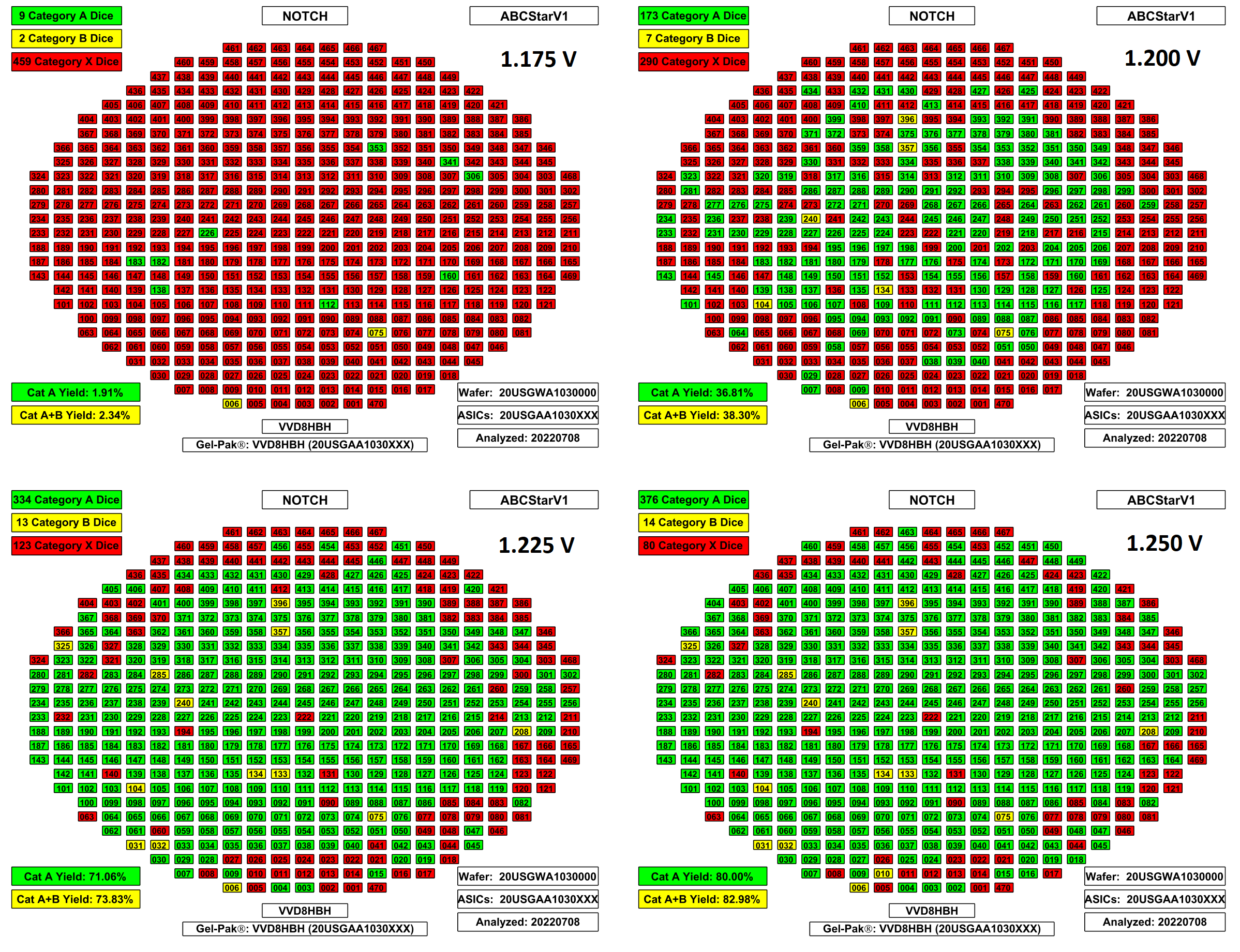}
\caption{\label{fig:vbadwafer} Digital core voltage dependency of a susceptible wafer}
\end{figure}

\subsubsection{Digital core voltage variation}
\label{sec:corevolt}

Varying the core voltage was the first avenue of exploration. Wafers with poor yields at the design
nominal digital core voltage of 1.20~V --- effectively the worst of the wafers in terms of transistor
speed and performance --- were tested to see what core voltage was required to achieve the targeted
Category A yield. It should be recalled here that tests that varied the analog core voltage during the
digital tests had no effect on the yields, and as such was not a factor that needed to be investigated
further. To illustrate the dramatic impact of digital core voltage changes, the wafer maps presented in
Figure \ref{fig:vbadwafer} shows the change in yield from nearly 2\% at 25~mV below the initial nominal
operating voltage of 1.20~V to 80\%+ at 50~mV above the nominal. Of particular importance for this wafer
is there was relatively little difference in yield between running it at 25~mV above nominal and 50~mV
above nominal. For ``good wafers'', there was often no difference in yield increasing the core voltage
to 1.250~V. This lack of, or minimal, difference in yields between a core voltage of 1.225~V and 1.250~V
was seen across all wafers that were tested, see for instance Table \ref{tab:coreyields}. This showed
that using a core voltage of 1.25~V for ABCStars in the ITk detector would provide a margin of
operational reliability. Given that the wafer testing was being done at room temperature, and the ASICs
would be run at an effective temperature --- depending on the chip's location --- of between
$-$16~$^\circ$C and $-$28~$^\circ$C (see \cite{itktdr}, Figure 9.10), the speed-ups would be cumulative
and provide further margin.

\begin{table}[htbp]
\centering
\caption{\label{tab:coreyields} Effect of increasing core voltage on yield across representative lots (Category A+B)}
\smallskip
\begin{tabular}{|c|c|c|c|c|c|}
\hline
Lot&Wafername&1.175~V&1.200~V&1.225~V&1.250~V\\
\hline
12YVB00000&VZDBAQH&56.38\%&82.55\%&87.23\%&87.23\%\\
\hline
144YB00000&VHDDLHH&79.57\%&85.96\%&86.17\%&86.17\%\\
144YB00000&VJDDLFH&65.11\%&79.79\%&81.70\%&81.70\%\\
\hline
183KB00000&VMDCEZH&88.30\%&88.72\%&88.72\%&88.72\%\\
183KB00000&VBDCDTH&90.43\%&91.70\%&91.70\%&91.70\%\\
\hline
15M2B00000&V7DDKAH&84.26\%&93.62\%&94.68\%&94.68\%\\
15M2B00000&VCDDJNH&79.57\%&90.21\%&91.49\%&91.49\%\\
\hline
2YEWB02000&V2DCCKH&57.02\%&84.89\%&88.72\%&88.94\%\\
2YEWB02000&V5DCFYH&67.45\%&92.77\%&94.04\%&94.04\%\\
\hline
0HZJB00000&VVD8HBH&2.34\%&38.30\%&73.83\%&82.98\%\\
0HZJB00000&VTD8LBH&2.34\%&38.09\%&74.26\%&80.43\%\\
\hline
\end{tabular}
\end{table}

Table \ref{tab:coreyields} shows how the core voltage required for dice to consistently pass varies
dramatically between lots. The primary contributor to dice failing differentially between different core
voltages was the A02 test. The first nine wafers in the table were taken from ``good lots'' that had
acceptable A02 yields at a 1.20~V digital core voltage, and the last two wafers in the table were from a
``bad lot'' with poor yields at that voltage. The average passing voltage across the first nine wafers
was 1.179~V, and the average passing voltage on the last two was 1.225~V.

\begin{figure}[htbp]
\centering 
\includegraphics[width=0.99\textwidth]{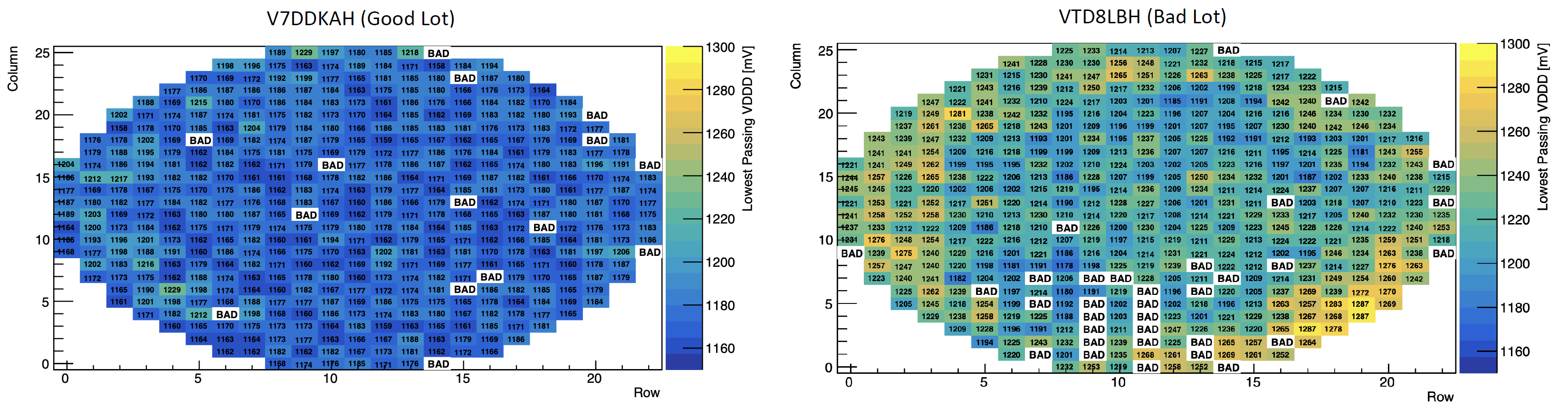}
\caption{\label{fig:a02passing} A02 passing core voltage --- good wafer vs. bad wafer}
\end{figure}

Another set of data visualizations produced was ``temperature maps'' for individual dice on wafers,
where the passing voltage of each chip was assigned a colour where ``cooler'' values were indicative of
lower passing voltages and thus chips with faster transistors, and ``hotter'' values for higher passing
voltages and slower transistors. An example of two wafers, one from a ``good lot'' and the other from a
``bad lot'' is shown in Figure \ref{fig:a02passing}. In the diagrams, white dice did not pass the
earlier tests in the suite and the A02 digital test could not be run on them. These would be Category X
(failed) dice regardless of the digital core voltage.

From the data gathered over the testing of numerous wafers at room temperature, it was determined that
setting the digital core voltage to 1.25~V for the ABCStar would offer a margin of operational safety in
the assembled detector, especially considering that the ASICs would be run much lower than room
temperature in the ITk. However, raising the core voltage from 1.20~V to 1.25~V would increase the digital
current to the chip from an average of 28.8~mA (34.6~mW) to 30.2~mA (37.8~mW), and going to 1.30~V would draw
an average of 31.7~mA (41.2~mW). Since the analog core voltage will continue to be 1.20~V in operation, and
consumes 68~mA of current (81.6~mW), increasing the digital core voltage to 1.25~V only results in an
overall increase in power dissipation of the ABCStar of roughly 2.8\%, and 5.7\% to go to 1.30~V. Given
that the initial design called for 233,856 ABCStar ASICs in the ITk (see \cite{itktdr}, Section 5.3
``Design of the Strip Modules''), going from 1.20~V to 1.25~V meant an overall additional 327~A to supply
and 748~W to dissipate, and going from 1.20~V to 1.30~V was an additional 678~A and 1,543~W. Thermoelectric
studies were performed that determined that raising the digital core voltage of the ABCStar to 1.30~V
would have minimal impact on the performance of the power and cooling infrastructure of the ITk as
already designed \cite{itktdr} --- and as such it would have a significant capability margin and would
not require modification to support the change to the lower 1.25~V.

\subsubsection{Duty cycle variation}
\label{sec:dutycycle}

\begin{figure}[htbp]
\centering 
\includegraphics[width=0.99\textwidth]{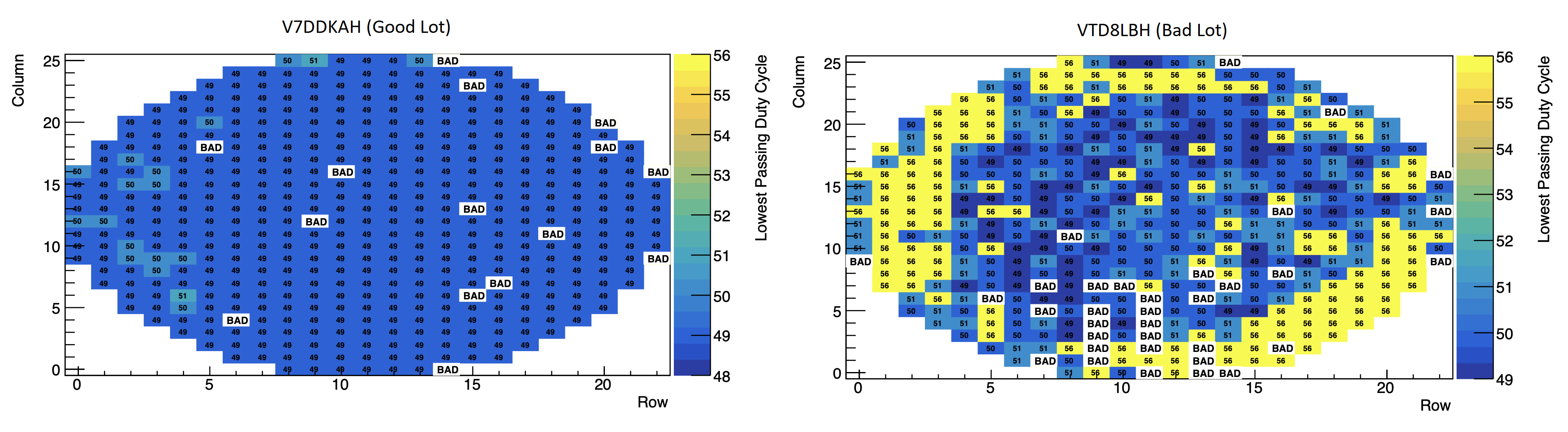}
\caption{\label{fig:a02dutycycle} A02 passing duty cycle at 1.20~V --- good wafer vs. bad wafer}
\end{figure}

On tests of wafers from various lots, it was determined that the duty cycle required for the A02 test to
pass at a digital core voltage of 1.20~V varied from 49\% to 56\% (where the duty cycle is the amount of
time the clock spent in a logic high state versus the logic low state), see Figure
\ref{fig:a02dutycycle}, recalling that this provided a longer settling time for the SRAM control
logic. This sensitivity to duty cycle was a source of concern even if the core voltage was raised to
1.25~V. The 40~MHz clock to the ABCStars is provided by HCCStar ASICs which allow for some tunability of
the duty cycle to compensate for manufacturing variation of that chip.

\begin{figure}[htbp]
\centering 
\includegraphics[width=0.99\textwidth]{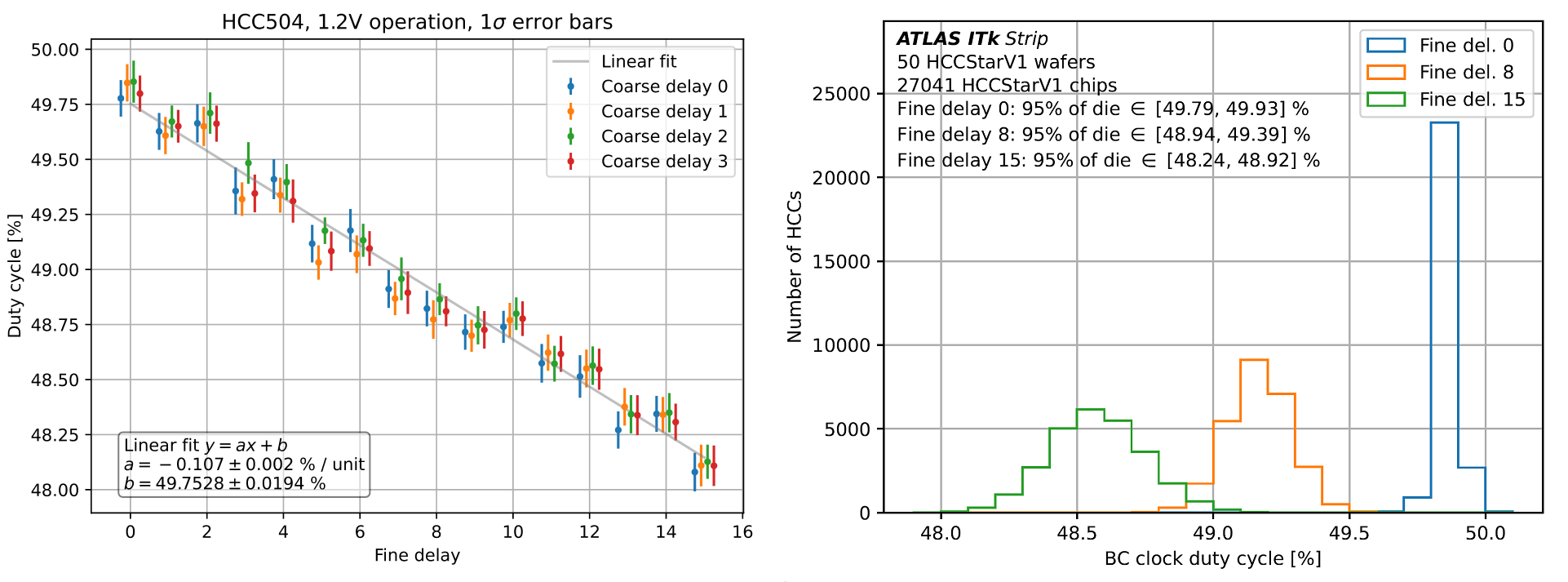}
\caption{\label{fig:hccstardc} HCCStar duty cycle tunability range for a representative chip and 50 production wafers}
\end{figure}

The HCCStar has two registers that control a coarse and fine delay of the 40~MHz clock as applied to the
ABCStars. Studies on the HCCStar showed that, even with the tuning parameters that affected duty cycle
at their maximum, the duty cycle provided to the ABCStar would always be less than 50\%, see Figure
\ref{fig:hccstardc}. Additionally, the studies also showed that at the other extreme of tunability, the
duty cycle could be reduced to as low as 48\%. This was obviously sub-optimal given the susceptibility
of the ABCStar to lower duty cycles and would need to be addressed.

The physical clock signal from the HCCStar to the ABCStar uses Scalable Low-Voltage Signalling (SLVS,
see for example \cite{slvs}), which consists of a pair of differential signaling lines that produce a
400~mV differential signal around a common-mode voltage nominally half the supply voltage of the chip:
one with a positive phase and the other with the opposite phase. Because the HCCStar generated a
positive clock duty cycle below 50\%, it was proposed that the positive and negative signals of the SLVS
differential pairs from the HCCStar to the ABCStars be physically swapped. If this was done, the duty
cycle of the ABCStar's clock could be guaranteed to be a minimum of 50\% and a maximum of 52\%, thus
providing a further margin of reliability to the operation of the ABCStar.

\subsubsection{Temperature variation and radiation effects}
\label{sec:radroach}

Studies were also done on a sampling of dice from various wafer lots mounted on test fixtures to verify
whether assumptions regarding transistor speed versus total ionizing radiation dose (TID) and
temperature were correct. Tests had previously been conducted on ABCStar chips to verify the design was
stable over all expected conditions, so these tests were primarily investigating the effect of radiation
and temperature on the digital core voltage that would guarantee a passing A02 test. As discussed in
Section \ref{sec:corevolt}, the increased digital core voltage and lower temperature in operation within
the ITk detector would provide a high margin of reliability for dice that could pass the tests at room
temperature during the wafer testing phase.

\begin{figure}[htbp]
\centering 
\includegraphics[width=0.9\textwidth]{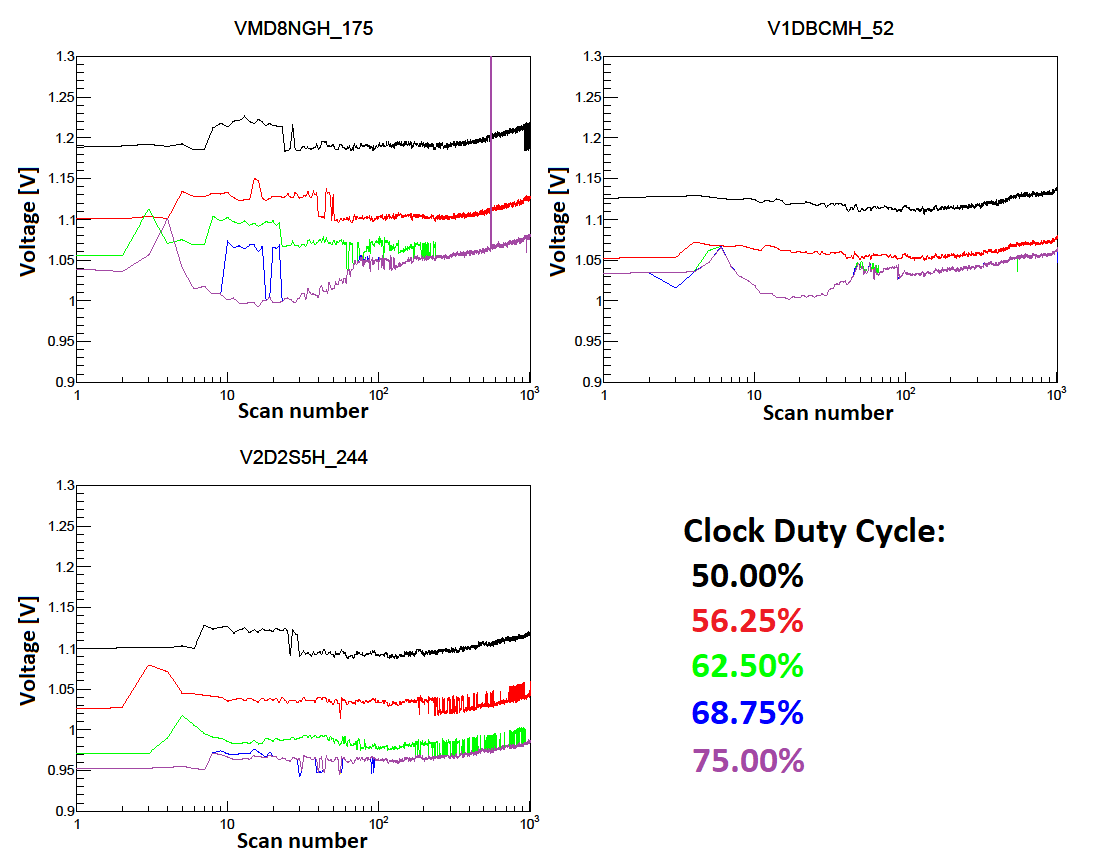}
\caption{\label{fig:radtestdc} Radiation effects on A02 passing core voltage on chips irradiated to 100~Mrad}
\end{figure}

Radiation tests were performed on a sampling of ABCStars from the original production lots. The test
chips were run at different clock duty cycles and the digital core voltage was lowered until unexpected
errors were seen in the output of the A02 test, which was run every 5 minutes while the ASICs received a
dose of 1~Mrad/hr. The test fixture consisted of a stack of four independent systems: each with a
commercial FPGA board, custom interface electronics, and a test board holding an ABCStar --- with only
the ABCStar boards in the beam path. Four ABCStars were irradiated to 25~Mrad, then four to 50~Mrad,
four to 75~Mrad, and three to 100~Mrad (one test setup experienced anomalies and did not provide useful
data). Results for the batch of chips that were irradiated up to 100~Mrad is shown in Figure
\ref{fig:radtestdc}. The results for the chips irradiated to lower TIDs are not shown as their
performance is similar to the measurements at lower radiation levels as those shown in Figure
\ref{fig:radtestdc}. Note: the increased supply voltage required between approximately the 10th and
100th scan is due to known radiation effects on the power requirements of the digital circuitry that
peak at around 1~Mrad before returning to pre-irradiation levels (see \cite{abc130}, section 4.10
``Current increase with Total Ionising Dose (TID bump)'' for details). It can be seen that, at a 50\%
duty cycle, the maximum required core voltage on the worst chip (wafer VMD8NGH, die 175) was roughly
1.225~V after the irradiation, up from roughly 1.195~V at lower radiation levels. This agreed with the
data from the other batches and showed that irradiated dice required a 20--30~mV increase in their
digital core voltage for reliable operation. The tests also further confirmed the improvements that
could be achieved through larger duty cycle variations (see Section \ref{sec:dutycycle}).

\begin{figure}[htbp]
\centering 
\includegraphics[width=0.9\textwidth]{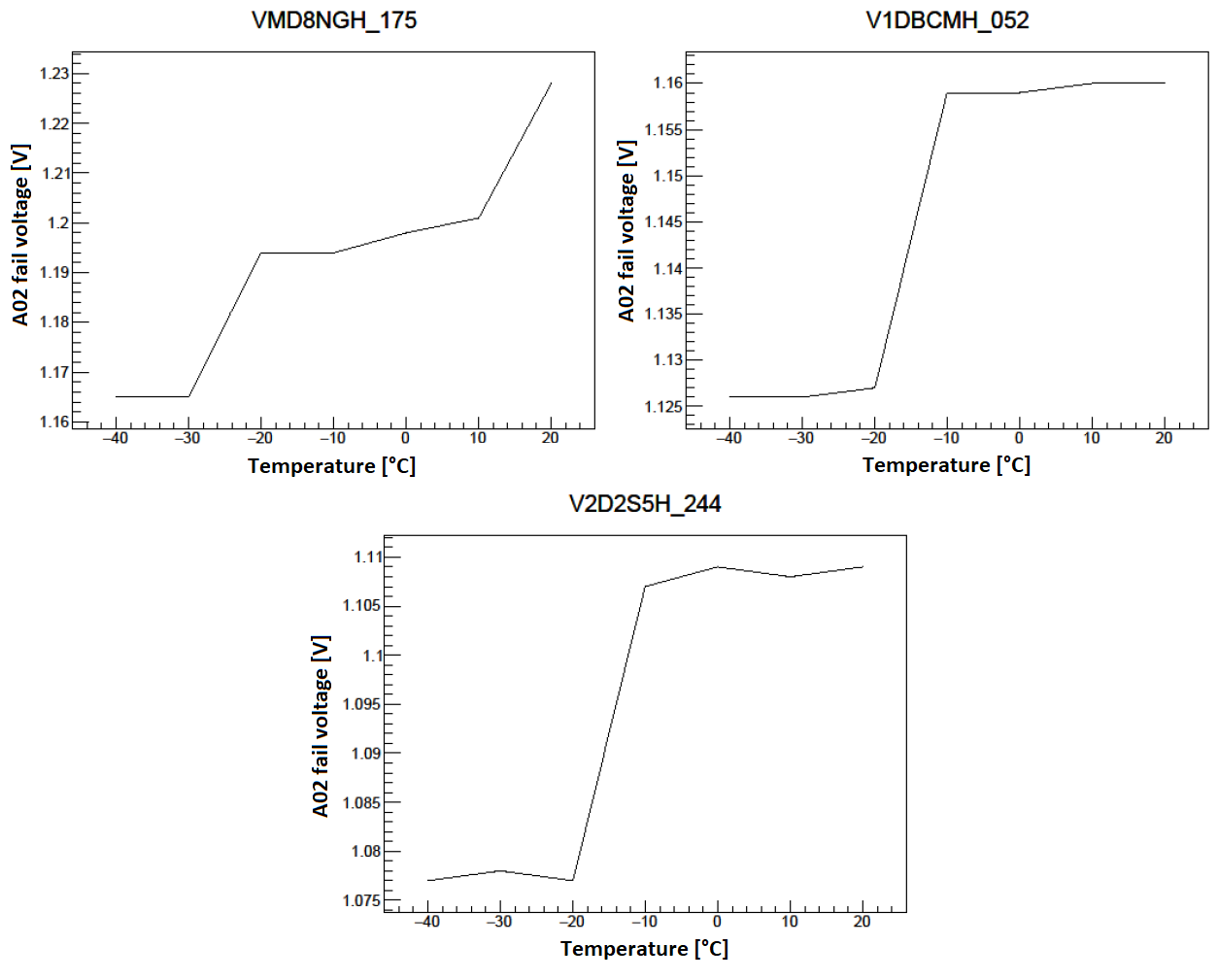}
\caption{\label{fig:temptesta02} Temperature effects on the passing A02 core voltage of irradiated ABCStars}
\end{figure}

After the irradiation, all of the batches of chips were placed together in a temperature chamber that
varied from $-$40~$^\circ$C to $+$20~$^\circ$C as the A02 test was run. All tests were done at a 50\%
clock duty cycle and a pass of the A02 test occurred only when there were zero errors in its output. The
temperature test results for the same chips shown in Figure \ref{fig:radtestdc} are shown in Figure
\ref{fig:temptesta02}. The temperature tests across ABCStars irradiated to different TIDs showed a
reduction in required digital core voltage levels of between 30~mV and 60~mV, which agrees with the
expected outcome of the experiment. Operating the ABCStar at the lower temperatures of the ITk will
improve the operating margin by speeding up its transistors.

\subsubsection{Other tests performed}
\label{sec:othertests}

Some studies were done with a decreased clock frequency for the ABCStar that showed this also mitigated
the SRAM issue. This test was only performed as a cross-check of the other results as the 40~MHz clock
applied to the ABCStar is the LHC beam crossing clock and cannot be altered. A further indirect
measurement of the speed of the transistors on any ABCStar can be accomplished using a built-in ring
oscillator circuit. The oscillator drives a free running counter register that can be read to determine
the oscillator's frequency. On chips with slower transistors, the oscillator will also run slower. It
was included on the ABCStar as a means of monitoring changes in transistor speed versus TID, but
provides an objective measurement at wafer test time as well. As such, the ring oscillator frequency can
also be used to qualitatively see the effects of various operational conditions on the transistor
speed. During production testing, the oscillator frequency was measured at room temperature and a
digital core voltage of 1.25~V and logged for later reference.

\subsection{Investigation of modified wafer production processes}
\label{sec:newprocess}

The main issue with the experimental wafer route was 300 of 700 planned wafers at the time had already
been manufactured with the standard process and could not be modified, 200 more had been started but
could still be modified, and any remaining new wafers and replacements for the existing ones could
potentially use a new process. Before committing to a new process, manufactured test wafers and diced
chips using the modified transistors would also have to go through extensive validation tests to make
sure there were no other unexpected side effects (e.g. susceptibility to radiation, temperature or power
supply changes, changes in analog performance, or timing related factors). Having to conduct the
required testing would have caused delays in the critical path for the assembly of the strips detector
modules. Given the expected costs and project delays this was not the preferred route, but was a
necessary one if the modifications to the operating conditions of the dice manufactured with the
standard process proved inadequate --- using one of the modified processes suggested by the foundry
might eliminate the need for a redesign and subsequent retooling and re-manufacturing of the ABCStar
ASICs.

\begin{figure}[htbp]
\centering 
\includegraphics[width=0.99\textwidth]{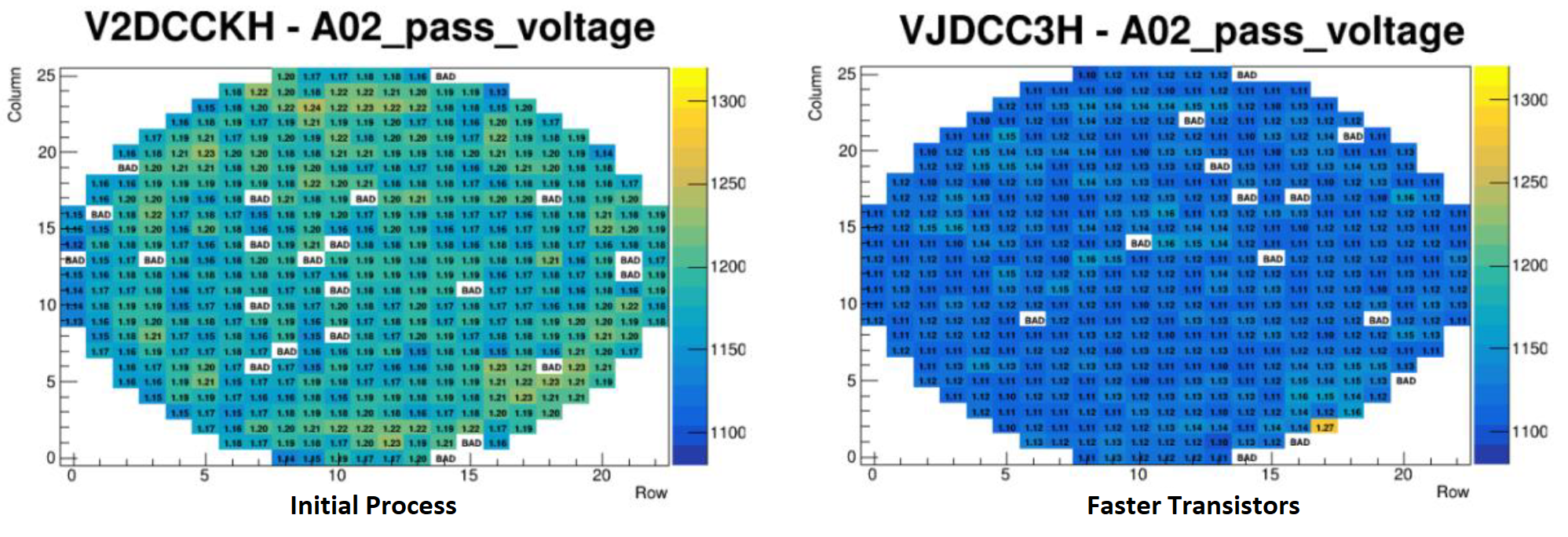}
\caption{\label{fig:expfast} Comparison of standard wafer (left) and ``fast transistor'' (right) A02 pass voltages}
\end{figure}

Three process changes were suggested by the foundry: reducing the gate size of the transistors by 2~nm or
4~nm, and modifying the process using a technique proprietary to the foundry known to speed up
transistors without modifying their geometry. Two experimental wafers each were produced using the three
different modifications, and a fourth pair of wafers were made with a combination of the 2~nm gate size
reduction and the faster transistors. As expected, the digital core voltage required for the A02 test to
pass was reduced on the experimental wafers, see for example Figure \ref{fig:expfast}. Table
\ref{tab:expa02v} presents more detailed results for the A02 passing voltage, with a 50\% clock duty
cycle, on a sampling of wafers manufactured with the standard process and all of the experimental
wafers. In the table, the $+$0 column is the die when it is tuned to the nominal design target core
voltage of 1.20~V, lower LDO Register settings produce higher core voltages, and higher values produce
lower voltages --- on average $\pm$27.8~mV per increment or decrement of the value.

\begin{table}[htbp]
\centering
\caption{\label{tab:expa02v} Summary of A02 core voltage sensivity in experimental vs. standard process wafers}
\smallskip
\begin{tabular}{|c|c|c|c|c|c|c|c|c|c|c|}
\hline
&&\multicolumn{7}{c|}{LDO Register Setting vs. Fail Count}&\multicolumn{2}{c|}{A02 Pass (mV)}\\
\cline{3-11}
Wafername&Process&$-$3&$-$2&$-$1&$+$0&$+$1&$+$2&$+$3&Mean&Sigma\\
\hline
V2DCCKH&standard&10&10&10&38&209&417&457&1183&22\\
V3DCCJH&standard&5&5&7&22&134&376&456&1174&21\\
V5DCFYH&standard&8&8&9&23&151&403&463&1177&19\\
V9DCCDH&standard&2&2&2&22&153&375&460&1176&21\\
\hline
VADCFTH&gate $-$2~nm&6&6&6&25&175&419&462&1180&19\\
VDDCC9H&gate $-$2~nm&6&6&6&7&54&286&442&1162&19\\
\hline
VHDCFLH&gate $-$4~nm&4&4&4&4&21&195&390&1152&20\\
VIDCFKH&gate $-$4~nm&8&8&7&7&24&168&382&1149&20\\
\hline
VIDCK0H&faster&4&4&4&4&9&72&285&1138&19\\
VJDCC3H&faster&1&2&2&2&2&5&67&1122&14\\
\hline
VJDCFJH&$-$2~nm + faster&7&7&7&7&7&16&120&1124&14\\
VUDCF9H&$-$2~nm + faster&8&8&8&9&9&21&107&1122&14\\
\hline
\end{tabular}
\end{table}

Further changes were also observed in the experimental wafers' analog circuitry performance that
required modification of wafer test result limits as the distribution of some measured values
shifted. In particular, all wafers made with the ``faster transistor'' process were observed to have
analog performance sufficiently different from the wafers made with the standard process that there was
concern about their usability. Specifically, the ``fast transistor'' wafers had a lower bandgap
reference voltage that required higher tuning values for the on-chip digital to analog converters,
resulted in higher gain in the front-end circuitry, reduced the available range for tuning the timing of
the digital portion of the chip, and gave higher voltages from the LDOs for a given register setting. In
looking at the test data, the 4~nm gate size reduction was seen as the best balance of improved digital
performance and having analog performance similar to the standard process wafers. Some of the analog
parameter changes are summarized in Table \ref{tab:expparms} and compared to wafers manufactured using
the standard process. In addition, the wafer test results (the number of dice in Category A, B, and X)
are presented both before and after the changes were made to the test limits applied to the measured
values. The wafers were not re-tested, but a different limits cut was applied to the data collected and
analysis results from the tests. In the table, ``BG'' is an on-chip bandgap reference that is amplified
and used for tuning DACs and other circuitry, and ``Osc.'' is the frequency of the on-chip 90~MHz nominal
ring oscillator at 1.20~V. As can be seen, the behaviour of this circuitry versus transistor speed
follows the trends that were expected with the process modifications.

\begin{table}[htbp]
\centering
\caption{\label{tab:expparms} Summary performance of experimental vs. standard process wafers}
\smallskip
\begin{tabular}{|c|c|c|c|c|c|c|c|c|c|c|}
\hline
&&\multicolumn{3}{c|}{Initial Cut}&\multicolumn{3}{c|}{Updated Cut}&BG&Gain&Osc.\\
\cline{3-8}
Wafername&Process&nA&nB&nX&nA&nB&nX&mV&mV/fC&MHz\\
\hline
V2DCCKH&standard&372&20&78&376&21&73&589&80.1&89.5\\
V3DCCJH&standard&327&28&115&361&44&65&593&81.4&90.1\\
V5DCFYH&standard&399&20&51&410&19&41&590&80.9&89.2\\
V9DCCDH&standard&325&32&113&371&40&59&592&82.2&89.0\\
\hline
VADCFTH&$-$2~nm&388&19&63&407&19&44&590&81.1&89.8\\
VDDCC9H&$-$2~nm&364&17&89&410&22&38&592&81.8&91.4\\
\hline
VHDCFLH&$-$4~nm&368&19&83&411&20&39&589&81.0&93.0\\
VIDCFKH&$-$4~nm&359&26&85&405&28&37&590&81.3&93.2\\
\hline
VIDCK0H&faster&300&17&153&423&19&28&575&83.3&93.3\\
VJDCC3H&faster&146&15&309&365&45&60&570&85.0&95.8\\
\hline
VJDCFJH&$-$2~nm + fast&227&11&232&378&22&70&575&83.1&95.4\\
VUDCF9H&$-$2~nm + fast&233&15&222&399&24&47&578&83.6&95.5\\
\hline
\end{tabular}
\end{table}

\section{Decisions for production and detector operation}
\label{sec:reliability}

The decisions on how to move into production of the ABCStar and the detector modules were based on the
observations gathered during the various mitigation studies described in Section \ref{sec:mitigation}
and summarized in Table \ref{tab:mitres}.

\begin{table}[htbp]
\centering
\caption{\label{tab:mitres} Summary of SRAM timing mitigation investigation results}
\smallskip
\begin{tabular}{|p{0.25\columnwidth}|p{0.23\columnwidth}|p{0.4\columnwidth}|}
\hline
Modification&Observation&Can we change this?\\
\hline
Increase digital supply voltage (LDO setting)&More dice pass tests, marked effect&Yes, effective and safe\\
\hline
Decrease clock frequency&More dice pass&No, 40~MHz set by LHC beam crossings\\
\hline
Faster transistors&More dice pass&Yes, the foundry can adjust wafers\\
\hline
Increase clock duty cycle provided by HCCStar&More dice pass at a higher duty cycle (high longer than
low), marked effect&No, the duty cycle out of the HCCStar is fixed at 49\%. However swapping positive
and negative signals of the diffential clock pair would give 51\%.\\
\hline
Lower temperature&More dice pass&We gain some benefit from ITk cooling, but there is a set cooling
profile through its operational lifetime (so not an available handle)\\
\hline
Higher radiation dose&Fewer dice pass, relatively small effect&No, the accumulated dose is set by
position in the detector and passage of time\\
\hline
\end{tabular}
\end{table}

The following decisions were made regarding the go-forward strategy for testing and production, which
has proven to be a robust framework to produce reliable ABCStars for the ITk detector module assemblies.

\begin{itemize}
\item For detector operation:
  \begin{itemize}
    \item Operate ABCStars in the detector using a higher setting for the on-chip digital LDO linear
      regulator, nominally tuned to 1.25~V.
    \item It will be possible, if needed, to develop a calibration routine for the detector to determine
      the lowest digital core voltage each individual ABCStar can be reliably operated at --- with the
      calibration re-run periodically during operation.
  \end{itemize}
\item For production probing:
  \begin{itemize}
    \item The digital tests would be run at three voltages: with the digital LDO tuned to 1.25~V, and then
      at the next higher two LDO register settings (which will generate lower voltages, $\sim$1.225~V and
      $\sim$1.20~V respectively). This procedure was intended to provide data on the lowest voltage step at
      which all digital tests pass. To pass, the tests all had to run at 1.25~V.
    \item The P99 test would be run at 100~mV below the new nominal operating voltage of 1.25~V, or 1.15~V. It
      was originally set to run at 1.10~V during pre-production.
  \end{itemize}
\item Because there was a mitigation strategy for the ABCStars manufactured with the standard process,
  it was decided not to pursue foundry adjustments for the remainder of the wafers being produced.
\item To provide a higher duty cycle, it was decided to swap the wirebonds between the HCCStar's
  differential clock's positive and negative signals and the printed circuit board --- the traces of
  which are routed to the ABCStars connected to it. This will guarantee that the duty cycle as provided
  by the HCCStar to the ABCStars will always be greater than 50\%.
\end{itemize}

\noindent All of the changes required were made to the production testing of the remaining ABCStar
wafers, which was completed in early 2025. The required changes have also been made in the specification
and implementation of the ITk control software and will be used during operation of the detector.

\section{Conclusion}
\label{sec:conclusion}

In this paper, technical responses to the discovery of a subtle logic timing error in the ABCStar, the
ITk strips detector front-end readout ASIC, were documented. The need for high reliability and
predictable performance over the operational lifetime of the ATLAS ITk detector was the primary driver
in considering solutions, with the possible need for a redesign of the chip being the worst case
scenario due to its impact on project schedules and costs. Two primary approaches were considered:
in-situ modification of the operating conditions for the ABCStar in the detector, and modifications to
the manufacturing process proposed by the wafer foundry to speed up transistors in the existing
design. Both approaches were pursued in parallel, including the manufacture and testing of eight
experimental wafers using several of the proposed process modifications. By understanding the root cause
of the timing issue and having an accurate understanding of the existing process, it was demonstrated
that an in-situ approach of raising the core voltage and increasing the duty cycle of the clock signal
to the chip would meet the requirements necessary for the reliable operation of the ITk detector. Wafer
yields of as low as 2\% were raised to over 80\% after the operating parameters were modified, and
allowed the production and testing of the ABCStar and subsequent detector module assembly and test to
proceed.

Because this is a key lesson learned, it needs to be emphasized that the decision to use the SRAM block
without a complete re-simulation was a conscious one not taken without cause. A major factor considered
was the design's long history in several ``silicon-proven'' implementations, including in other ASICs,
and there was every confidence in the demonstrable legacy of the SRAM block as designed. The use of such
blocks with the assumption they are correct is standard practice in industry to provide reduced
development time and effort, and reduce risk. While the decision to include the block without deeper
analysis is the source of the failures that were seen during testing, it was an issue that was exposed
by manufacturing process variation rather than it being an inherent logic flaw. In fact, further testing
confirmed the design soundness of the SRAM block's logic when the process variation was compensated for
through the digital core voltage increase and would provide reliable operation over the lifetime of the
ITk dectector. It is instructive that the success of a design block incorporated into the ABCStar was
the source of an informed decision that ultimately proved ill-advised, and this is a lesson that is
generally applicable to the design of any integrated circuit, and ASICs in particular.

Another lesson learned that should be applied to future ASIC designs for detectors and other projects
--- to mitigate the risks associated with the use of presumed-reliable ``silicon proven'' circuit blocks
--- is to request that the foundry produce a set of what are called ``striped'' wafers that have dice on
them that are manufactured with all extremes of expected process variation on the same wafer. If this
had been done with the ABCStar, the process dependent timing issue would have been caught on the
``stripes'' with slower transistors. This technique is available for many foundry processes, and
provides one last chance to catch marginal designs before they go into pre-production or production
testing. If there are other potential process-related yield issues, this technique could potentially
catch those as well and steps could be taken to do a redesign if needed before full production
begins. If this is not possible for a project, early testing should be done at reduced
voltages to simulate process variation and determine the margins of reliable operation of the first
wafers produced.

With the integration of the ITk detector beginning in late 2025, and final commissioning scheduled for
2028, overcoming the challenge with the ABCStar design was critical in meeting those project goals. It
is hoped that the experiences shared here will be of use in informing other efforts to ensure the
reliable operation of integrated circuits used in detectors, and other systems, that require
high-reliability in hostile operating environments.

%

\acknowledgments

We are grateful to the management and staff of DA-Integrated for their help in identifying and implementing
the mitigating solutions to the ABCStar readout ASIC. This work was supported by the Canada Foundation for
Innovation, the Natural Science and Engineering Research Council of Canada, and the Department of Energy of
the United States.



\end{document}